\documentclass[pra,twocolumn,aps,showpacs,superscriptaddress]{revtex4-1}
\usepackage{amsmath}
\usepackage{amssymb}
\usepackage{hyperref}
\usepackage{graphicx}
\usepackage{color}

\def\B{{\rm B}}
\def\F{{\rm F}}
\def\BF{{\rm BF}}
\def\FB{{\rm FB}}
\def\BB{{\rm BB}}
\def\FF{{\rm FF}}

\begin{document}

%%%%%%%%%%%%%%%%%%%%%%%%%%%%%%%%%%%%%%%%%%%%%%%%%%%%%%%%%%%%%%%%%%%%%%%%%%%%%%%
%%%%                     Title and authors                                 %%%%
%%%%%%%%%%%%%%%%%%%%%%%%%%%%%%%%%%%%%%%%%%%%%%%%%%%%%%%%%%%%%%%%%%%%%%%%%%%%%%%

\title{Position swapping and pinching in  Bose-Fermi mixtures with two-color 
       optical Feshbach resonances}

\author{S. Gautam} 
\affiliation{Physical Research Laboratory,
             Navarangpura, Ahmedabad - 380 009, India}
\author{P. Muruganandam} 
\affiliation{ Center for Nonlinear Dynamics, Department of Physics, 
              Bharathidasan University, Tiruchirapalli 620 024, 
              Tamil Nadu, India}
\author{D. Angom}
\affiliation{Physical Research Laboratory,
         Navarangpura, Ahmedabad - 380 009, India}

%%%%%%%%%%%%%%%%%%%%%%%%%%%%%%%%%%%%%%%%%%%%%%%%%%%%%%%%%%%%%%%%%%%%%%%%%%%%%%%
%%%%%%%%%%%%%                   Abstract                           %%%%%%%%%%%%
%%%%%%%%%%%%%%%%%%%%%%%%%%%%%%%%%%%%%%%%%%%%%%%%%%%%%%%%%%%%%%%%%%%%%%%%%%%%%%%

\date{\today}
\begin{abstract}
We examine the density profiles of the quantum degenerate Bose-Fermi mixture
of $^{174}$Yb-$^{173}$Yb, experimental observed recently, in the mean field 
regime. In this mixture there is a possibility of tuning the Bose-Bose and 
Bose-Fermi interactions simultaneously using two well separated optical 
Feshbach resonances, and it is a good candidate to explore phase separation 
in Bose-Fermi mixtures. Depending on the Bose-Bose scattering length $a_\BB$,
as the Bose-Fermi interaction is tuned the density of the fermions is 
pinched or swapping with bosons occurs.
\end{abstract}


%% 03.75.Hh : Static properties of condensates; thermodynamical, 
%%            statistical, and structural properties 
%% 67.85.Pq : Mixtures of Bose and Fermi gases

\maketitle

%%%%%%%%%%%%%%%%%%%%%%%%%%%%%%%%%%%%%%%%%%%%%%%%%%%%%%%%%%%%%%%%%%%%%%%%%%%%%%%
%%%%%%%%%%%%%                 Introduction                         %%%%%%%%%%%%
%%%%%%%%%%%%%%%%%%%%%%%%%%%%%%%%%%%%%%%%%%%%%%%%%%%%%%%%%%%%%%%%%%%%%%%%%%%%%%%

\section{Introduction}
The quantum degeneracy in a Boson-Fermi mixture was first experimentally 
realized for the system consisting of $^7$Li and $^6$Li 
\cite{Truscott,Schreck}. Since then it has been observed in several other 
Bose-Fermi mixtures, $^{23}$Na-$^{6}$Li \cite{Hadzibabic}, 
$^{87}$Rb-$^{40}$K \cite{Roati}, $^{87}$Rb-$^{6}$Li \cite{Silber}, 
$^4$He-$^3$He\cite{McNamara}, $^{174}$Yb-$^{173}$Yb\cite{Fukuhara}, and 
$^{84}$Sr-$^{87}$Sr \cite{Tey}. These are candidate systems to explore the
effects of boson-induced fermionic interactions, of particular interest is the
boson mediated fermionic superfluidity. Another property of interest 
is the dynamical instabilities of the fermionic component arising from the 
attractive fermionic interactions, which is also boson mediated \cite{Modugno}. 
Precondition to observe either of these is a precise control of the inter 
species interaction through a Feshbach resonance. Which has been observed in 
$^{87}$Rb-$^{40}$K \cite{Ferlaino-06,Zaccanti} and used to trigger the 
dynamical collapse of $^{40}$K \cite{Modugno}, the same is  numerically 
analyzed in ref.\cite{Modugno-03,Adhikari}. 

  Density distributions in the phase separated domain of the Bose-Fermi mixture 
is also an important property of interest. Like in binary condensates, 
dynamical instabilities can be initiated in the phase separated domain through
manipulations of interaction strengths. For binary condensates, the recent work
on mixtures of two different hyperfine states of $^{87}$Rb \cite{Tojo} is a 
fine example of controlled experiment on phase separation. 

In this regard, Molmer and collaborator \cite{Molmer, Nygaard} examined the 
zero temperature equilibrium density distributions and predicted widely 
varying density patterns as a function of inter species interactions. However, 
the bose-bose and bose-fermi interactions considered are extremely strong for 
experimental realizations. Similar studies have examined the ground state 
geometry in spherical traps \cite{Minguzzi,Roth}. The conditions for 
mixing-demixing have been analyzed for homogeneous Bose-Fermi mixture 
\cite{Viverit} and Bose-Fermi mixtures inside traps \cite{Minguzzi,Akdeniz}. 
Although, very high interaction strengths are achievable through magnetic 
Feshbach resonances in alkali atoms, simultaneous tuning of both the 
boson-boson and boson-fermion interactions is not possible. However, 
simultaneous tuning is possible with optical Feshbach resonances (OFR) when 
the resonant frequencies of the boson-boson and boson-fermion interactions are 
well separated. 

  With the realization of quantum degeneracy in the mixture of 
$^{174}$Yb-$^{173}$Yb where intra-species interactions for $^{174}$Yb can be 
tuned by OFR \cite{Yamazaki}, we find it pertinent to 
revisit these studies. With the possibility of tuning inter-species 
interactions for $^{174}$Yb-$^{173}$Yb mixture, it may be possible to realize 
the ground state geometries which are hitherto elusive. We, therefore, 
consider $^{173}$Yb-$^{174}$Yb mixture to study the density profiles for various
values of coupling strengths in the present work. It must also be mentioned
that the isotopes of Yb exhibit a wide range of inter- and intra-species 
interactions. It has attracted lot of attention as selected isotopic 
compositions  may exhibit dynamical instabilities triggered through the
interactions. The  $^{174}$Yb-$^{176}$Yb is one such Bose-Bose binary mixture 
currently investigated for instabilities on account of the attractive 
intra-species interaction of $^{176}$Yb \cite{Kasamatsu,Chaudhury}.

 The paper is organized into four sections. In the next section, Section.II,
we provide a brief description of the mean field equations of bosons and 
fermions. This is followed with the section on phase separation, where the
nature of bose-fermi phase separation is discussed as a function of the 
inter-species interaction. More importantly, the occurance of fermion pinching
is explored. Position swapping between bosons and fermions, as the 
inter-species interaction is increased, is then examined in the next section. 
We then conclude  with Section. V.

%%%%%%%%%%%%%%%%%%%%%%%%%%%%%%%%%%%%%%%%%%%%%%%%%%%%%%%%%%%%%%%%%%%%%%%%%%%%%%%
%%%%              Zero temperature mean field description                 %%%%%
%%%%%%%%%%%%%%%%%%%%%%%%%%%%%%%%%%%%%%%%%%%%%%%%%%%%%%%%%%%%%%%%%%%%%%%%%%%%%%%

\section{Zero temperature mean field description}

We examine the stationary state properties of a Bose-Fermi (BF) mixture
consisting of $^{174}$Yb and $^{173}$Yb in spherically 
symmetric trapping potentials
\begin{align}
  V_\B(\mathbf{r}) = \frac{1}{2} m_\B \omega^2 r^2,\;\;
  V_\F(\mathbf{r}) = \frac{1}{2}m_\F\omega^2 r^2
\label{eq.pots}
\end{align}
where the subscripts $\B$ and $\F$ stand for boson and fermion, respectively, 
and $\omega$ is the radial trap frequency for the two components. The fermions 
are spin-polarized ( single species ) and the fermion-fermion mean field 
interactions arise from the degeneracy pressure \cite{Butts}. Whereas, 
the boson-boson and boson-fermion interactions arise from the $s$-wave 
scattering between the atoms. Considering these, the mean field 
energy functional of the Bose-Fermi mixture is
\cite{Capuzzi}
\begin{eqnarray}
  E[\Psi_\B,\Psi_\F] & = &\int d\mathbf{r}\biggl [ N_{\B}\left ( \frac{\hbar ^2 
     |\nabla \Psi_\B|^2}{2m_B}+V_\B|\Psi_\B|^2 \right . \biggr. 
                   \nonumber \\
  & &\left. +\frac{1}{2}G_{\BB}|\Psi_\B|^4\right ) 
     + N_{\F}\left ( \frac{\hbar^2 |\nabla \Psi_\B|^2}{6m_\F}
     +V_\F|\Psi_\B|^2 \right. \nonumber \\
  & &\biggl. \left. +\frac{3}{5}A|\Psi_\F|^{10/3}\right)
   + G_{\B\F}N_{\B}|\Psi_\B|^2|\Psi_F|^2\biggr ] ,
\end{eqnarray}
where $\Psi_\B(\mathbf{r},t)$ and $\Psi_\F(\mathbf{r},t)$ are bosonic and 
fermionic wave functions satisfying the normalization condition
\begin{equation}
\int d\mathbf{r}|\Psi_\B(\mathbf{r},t)| =
\int d\mathbf{r}|\Psi_\F(\mathbf{r},t)| = 1.
\end{equation}
Here $G_{\BB} = 4\pi \hbar^2 a_{\BB} N_{\B}/m_\B$, where $a_{\BB}$ is
the bosonic $s$-wave scattering length and $N_\B$ is the number of bosons,
is the bosonic intra-species interaction,
$G_\BF = 2\pi \hbar^2 a_\BF N_\F/m_R$ and
$G_\FB = 2\pi \hbar^2 a_\BF N_\B/m_R$,
where $m_R = (m_\B m_\F)/(m_\B + m_\F)$ is the reduced mass, $N_\F$ is the
number of fermions, and $a_{\BF}$ is the inter-species scattering length,
are the inter-species interactions, and 
$A=\hbar^2(6\pi^2N_\F)^{2/3}/(2m_\F)$. The Lagrangian of 
the system
\begin{equation}
   L = \int d{\bf r}\frac{i\hbar}{2}\sum_{i=\B,\F}\left(\Psi_i^{\star}
       \frac{\partial \Psi_i}{\partial t} - \Psi_i\frac{\partial 
       \Psi_i^{\star}}{\partial t}\right) -E[\Psi_\B,\Psi_\F].
\end{equation}
Using the action principle
\begin{equation}
\delta\int_{t_1}^{t_2} L dt = 0,
\end{equation}
we get a set of coupled partial differential equations
\begin{subequations}
  \begin{align}
     i\hbar \frac{\partial \Psi_\B}{\partial t} =   & \, 
     \biggr[- \frac{\hbar^2 \nabla^2}{2m_\B}  + V_\B(\mathbf{r}) + 
     G_{\BB} \vert \Psi_\B \vert^2 \notag \\
     &\, + G_{\BF} \vert \Psi_\F \vert^2 \biggr] \Psi_\B , \\
     i\hbar \frac{\partial \Psi_\F}{\partial t} =  & \, \biggr[- 
     \frac{\hbar^2 \nabla^2}{6m_\F}  + V_\F(\mathbf{r}) + 
     A \vert \Psi_\F \vert^{4/3} \notag \\
     &\, + G_{\FB} \vert \Psi_\B \vert^2 \biggr] \Psi_\F.
  \end{align}
  \label{eq:gpbf}
\end{subequations}

It is more convenient to rewrite Eqs.~(\ref{eq:gpbf}) in a dimensionless 
form by defining dimensionless parameters in terms of the frequency 
$\omega$ and the oscillator length $a_{\text{ho}} = \sqrt{\hbar/(m_\B
\omega)}$. Using ${\tilde {\mathbf{ r}}}= \mathbf{ r}/a_{\text{ho}}$, 
$\tilde t = t \omega$ as the scaled dimensionless variables of 
length and time, respectively, Eqs.~(\ref{eq:gpbf}) can be  rewritten as
\begin{subequations}\label{eq:gpbf_full}
  \begin{align}
     i\frac{\partial \psi_\B}{\partial \tilde t} = & \,\biggl[ -\frac{
        \nabla_{\tilde{\mathbf{r}}}^2}{2} + V_\B(\tilde{\mathbf{r}}) + g_{\BB} 
        \left\vert \psi_\B \right\vert^2 \notag \\
     &\, + g_{\BF} \left\vert \psi_\F \right\vert^2 \biggr]\psi_\B, \\
     i\frac{\partial \psi_\F}{\partial \tilde{t}} = &\, \biggr[ - \frac{
        \nabla_{ \tilde{\bf r}}^2}{6m_\text{ratio}} + m_{\text{ratio}} 
        V_\F(\tilde{\bf r})  +  g_{\FF} \left\vert \psi_\F \right\vert^{4/3} 
        \notag \\ 
     &\, + g_{\FB} \left\vert \psi_\B \right\vert^2 \biggr] \psi_\F
  \end{align}
\end{subequations}
where the rescaled wave functions are 
$\psi_\B = a_{\text{ho}}^{3/2} \Psi_\B(\tilde {\bf r}, \tilde t)$ and 
$\psi_\F = a_{\text{ho}}^{3/2} \Psi_\F(\tilde {\bf r}, \tilde t)$. Similarly,
the interaction strength parameters are
\begin{align}
   & g_{\BB} = \frac{4\pi a_{\BB} N_{\B}}{a_{\text{ho}}},\;\;\;
     g_{\BF} = \frac{2\pi a_{\BF} N_{\F}}{m_R\, a_{\text{ho}}},\notag \\
   & g_{\FF} = \frac{(6\pi^2N_\F)^{2/3}}{2m_\text{ratio}},\;\;\;
     g_{\FB} = \frac{2\pi a_{\BF} N_\B}{m_R \, a_{\text{ho}}}, \notag
\end{align}
with $m_\text{ratio} = m_\F/m_\B $. For the sake of simplicity, we represent 
the scaled quantities without the tilde ( $\tilde{}$ ) in the rest of the 
article. For spherically symmetric trapping potential, the Eqs(\ref{eq:gpbf}) 
are reduced to one dimensional coupled mean field equations
\begin{subequations}\label{eq:gpbf_r}
  \begin{align}
    i\frac{\partial \psi_\B}{\partial t} = & \,\left[ -\frac{1}{2}
\frac{\partial^2}{\partial r^2} + \frac{r^2}{2} + g_{\BB}\left\vert 
\frac{\psi_\B}{r} \right\vert^2 + g_{\BF} \left\vert \frac{\psi_\F}{r} 
\right\vert^2 \right]\psi_\B, \\
i\frac{\partial \psi_\F}{\partial t} = &\, \Biggr[ -\left(  
\frac{1}{3m_{\text{ratio}}}\right) \frac{1}{2}\frac{\partial^2}
{\partial r^2} + m_{\text{ratio}} \frac{r^2}{2}  \notag \\
&\, +  g_{\FF} \left\vert \frac{\psi_\F}{r} \right\vert^{4/3} + g_{\FB} 
\left\vert \frac{\psi_\B}{r} \right\vert^2 \Biggr] \psi_\F .
\end{align}
\end{subequations}
These are the coupled mean field equations which describe the bose-fermi 
mixture in trapping potentials at zero temperature. 
To obtain the stationary solutions, we solve the equations numerically 
using Crank-Nicholson scheme \cite{Muruganandam} with imaginary time 
propagation.

%%%%%%%%%%%%%%%%%%%%%%%%%%%%%%%%%%%%%%%%%%%%%%%%%%%%%%%%%%%%%%%%%%%%%%%%%%%%%%%
%%%%            Section III: Stationary state structure                   %%%%%
%%%%%%%%%%%%%%%%%%%%%%%%%%%%%%%%%%%%%%%%%%%%%%%%%%%%%%%%%%%%%%%%%%%%%%%%%%%%%%%

\section{Phase separation}

  Broadly speaking, for large values of $G_{\BB}$, the density profiles of the 
boson-fermion mixture in spherical symmetric traps can have three distinct 
geometries in phase separated regime: (a) fermionic core surrounded by bosonic 
shell, (b) bosonic core surrounded by fermionic shell, and (c) shell of bosons
between fermionic core and fermionic outer shell \cite{Molmer}. The 
inter-species interactions begin to play an important role, in determining
the stationary state structure, when the density profiles of the bosons and 
fermions are of similar spatial extent. This occurs when the bosonic 
intra-species interaction is strong. In the Thomas-Fermi (TF) approximation, 
the necessary condition for a mixture of equal number of bosons and fermions
($N_\B=N_\F=N $ ) is \cite{Molmer}
\begin{equation}
  a_\BB\approx 1.68 m_{\rm ratio}^{-5/2}a_{\rm ho}N^{-1/6}.
\end{equation} 
Here after, we use $ a^*_\BB$ to represent the particular value of $a_\BB $ at 
which bosons and fermions have the same spatial extent.For species which are 
isotopes of the same element the mass difference is small and 
$m_{\rm ratio}\approx 1$. The condition is then reduced to 
\begin{equation}
  a^*_\BB\approx 1.68 a_{\rm ho}N^{-1/6}.
 \label{eqn_neql}
\end{equation}
Considering $N\sim 10^6 $, which is the typical value in experimental 
realizations, $a^*_\BB \approx 0.17a_{\rm ho}$. As $a_{\rm ho}$ is in 
general $\sim 10^{-6}$m for harmonic trapping potentials, the required
value of $a_\BB$ is in the strongly interacting domain. It could be achieved 
when the interaction is tuned through a Feshbach resonance, magnetic in the
case of alkali-Earth metal atoms. With the overlapping density profiles, 
more intricate density patterns are observed when the inter-species 
interaction is increased, however, tuning $a_\BF$ with magnetic Feshbach 
resonance is ruled out. This complication does not arise when the interactions 
are tuned with well separated OFRs. In which case, the isotopes of two-valence 
lanthanide atom Yb is a suitable candidate. It has seven stable isotopes: five 
bosons ($^{168}$Yb, $^{170}$Yb, $^{172}$Yb, $^{174}$Yb, and $^{176}$Yb) and 
two fermions($^{171}$Yb, and $^{173}$Yb ), and homo-nuclear OFRs of bosonic 
isotopes ($^{172}$Yb and $^{176}$Yb ) were recently studied \cite{Enomoto}. 
Among the various possible species pairings $^{174}$Yb-$^{173}$Yb, which has 
positive intra- and inter-species background scattering lengths, is an ideal 
candidate to study bose-fermi mixtures in the strongly interacting domain. 

 For our studies, we consider a $^{174}$Yb-$^{173}$Yb mixture 
containing $10^6$ atoms of each species and trapped by spherically symmetric
trap with trapping frequency $\omega/(2\pi)=400$Hz. The $a_\BB$ is chosen to be 
equal to $1100a_0$, which is achievable with the OFR of 
the $6s^2\; ^1S_0\rightarrow 6s6p\; ^3P_1$ inter-combination transition. And,
the bose-fermi inter-species scattering length $a_\BF$ can be tuned with 
OFR of the allowed $6s^2\; ^1S_0\rightarrow 6s6p\; ^1P_1$ transition. This is 
a broad line and the disadvantage of using it is high atom loss rate. However, 
a major advantage of OFR tuned interactions is the fine spatial control it 
provides. Recently, submicron modulation of scattering length using Feshback 
resonances was achieved in $^{174}$Yb \cite{Yamazaki}. Such precise control on 
the spatial variation of interaction strength is unrealistic with magnetic 
Feshbach resonances. From here on, where it is not explicitly mentioned, 
reference to bose-fermi mixture implies $^{174}$Yb-$^{173}$Yb isotope mixture. 
To examine the density profiles of the mixture in the strongly interacting 
domain, we keep $a_\BB$ fixed and vary $a_\BF$ so that the system progresses 
from mixing to full demixing regime via partial demixing regime.
It must be emphasized that with TF approximation, from Eq. (\ref{eqn_neql})
the spatial extent of density profiles with $a_{BF}=0$ are same when 
$a^*_\BB = 1191.71a_o$. However, we have chosen $a_\BB = 1100a_o $, 
approximately the value at which the density profiles begin to exhibit the 
features of interest.

%%%%%%%%%%%%%%%%%%%%%%%%%%%%%%%%%%%%%%%%%%%%%%%%%%%%%%%%%%%%%%%%%%%%%%%%%%%%%%%
%%%%          Subsction III.A: Mixing to partial demixing regime          %%%%%
%%%%%%%%%%%%%%%%%%%%%%%%%%%%%%%%%%%%%%%%%%%%%%%%%%%%%%%%%%%%%%%%%%%%%%%%%%%%%%%

\subsection{Mixing to partial demixing regime}

  Starting from the initial conditions of the mixture, which as mentioned 
earlier is equal spatial extent of the component species and $a_\BF\approx0$, 
the value of $a_\BF$ is increased. To analyze the evolution of density
profiles, consider the TF profile of the boson and fermions in scaled units 
as defined earlier
\begin{subequations}
  \begin{align}
  n_\F(r) = & \frac{1}{6\pi ^2}\left\{ 2m_{\rm ratio}
              \left [ E_\F - V_\F(r) - u_\FB n_\B(r)\right ] \right\} ^{3/2}, \\
  n_\B(r) = &\frac{1}{u_\BB}\left [ \mu - V_\B(r) - u_\BF n_\F(r) \right ],
  \end{align}
 \label{eqn_n_r}
\end{subequations}
where $u_{\rm XY}=g_{\rm XY}/N_{\rm Y}$ and $E_\F$ is the fermi energy. From 
these expressions, the densities at the origin in absence of
inter-species interaction are
\begin{subequations}
  \begin{align}
  n_\F(0) = & \frac{1}{6\pi ^2}\left ( 2m_{\rm ratio} E_\F \right )^{3/2}, \\
  n_\B(0) = & \frac{\mu}{u_\BB}.
  \end{align}
 \label{eqn_n_0}
\end{subequations}
\begin{figure}[h]
  \includegraphics[width=8.5cm]{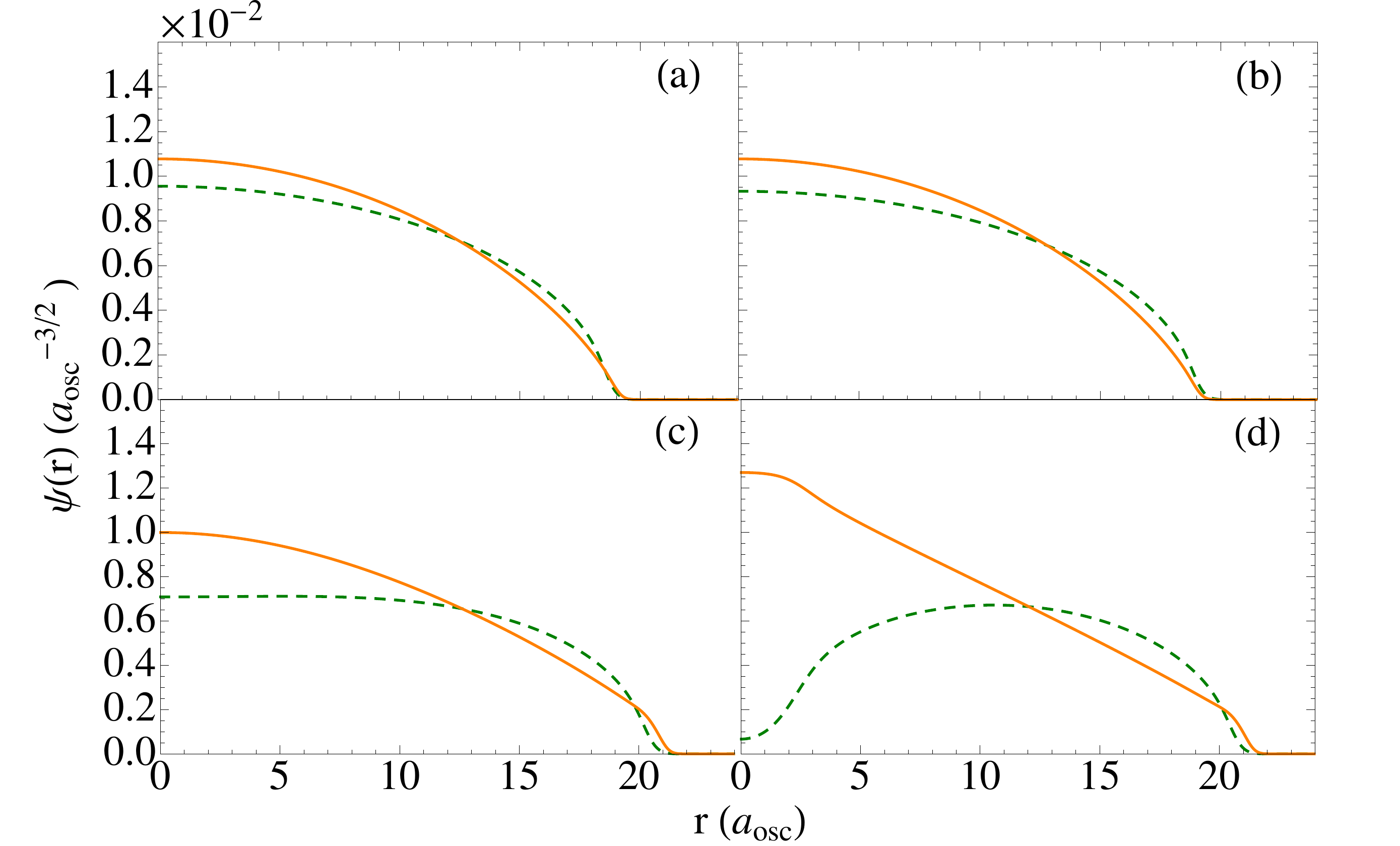}
  \caption{Wave-function profiles of $^{174}$Yb (dashed-green line) and
           $^{173}$Yb (solid-orange line) in a mixture of
            $^{174}$Yb-$^{173}$Yb with $N_\B=N_\F=10^6$, 
           $\omega/(2\pi)=400$Hz and $a_\BB=1100a_0$ as $a_\BF$ is changed
           from mixing to partial demixing domain. (a) For $a_\BF=0.0$, 
           (b) for $a_\BF=0.0a_\BB$ and $a_\BB=1197.11a_0$,
           (c) for $a_\BF=0.6a_\BB$, and (d) for $a_\BF=0.7a_\BB$}
  \label{part_demixing}
\end{figure}

In TF approximation, the fermi energy and chemical potential of the two species
are
\begin{subequations}
  \begin{align}
  E_\F = & (6N)^{1/3}, \\
  \mu  = & \frac{1}{2} \left (15a_\BB N\right)^{2/5}.
  \end{align}
 \label{eqn_ef_mu}
\end{subequations}
Using these in Eq.(\ref{eqn_n_0}), the ratio of the densities at the 
origin is
\begin{equation}
  \frac{n_\F(0)}{n_\B(0)} = 0.76a_{\BB}^{3/5}(15N)^{1/10}.\nonumber
\end{equation}
Consider $a_\BB\approx 0.17$, the value at which the profiles of the two 
species match for  $N=10^6$. The population ratio at the origin is then
\begin{equation}
  \frac{n_\F(0)}{n_\B(0)}  \approx 1.35.
\end{equation}
That is, when $a_\BF=0$ the fermion density is higher than the boson density
at the center of the trap. Given this as the initial condition, when the 
inter-species interaction is switched on, the inter-species mean field energy 
$G_\BF|\Psi_\F(0)|^2 > G_\FB|\Psi_\B(0)|^2$. Hence, it is energetically
favorable to shift the bosons from the center towards the edge of the trap. 
This is evident in the numerically obtained density profiles shown in 
Fig. \ref{part_demixing}(a), where the density profile of the bosons is 
flattened around the origin and has higher density at the edges. 

\begin{figure}[h]
   \includegraphics[width=8.5cm]{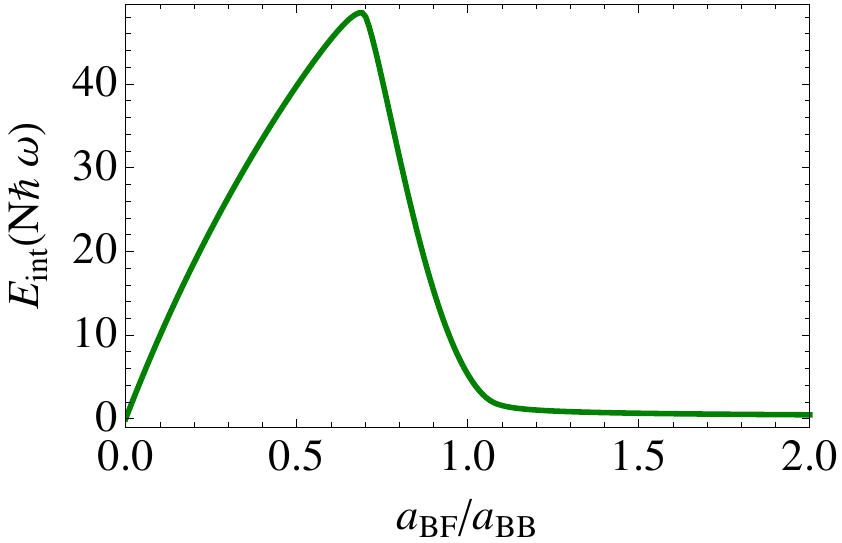}
   \caption{Interspecies interaction energy $E_{\rm int}$ between $^{174}$Yb 
            (boson) and $^{173}$Yb (fermion) as a function of $a_\BF$. The 
            maxima of $E_{\rm int}$ occurs at $a_\BF=0.7a_\BB$.}
   \label{eint_bf}
\end{figure}

The inter-species interaction energy in mixing regime is
\begin{eqnarray}
E_{\rm int}&=&\int d\mathbf{r}u_{\B\F}n_{\B}n_{\F},\nonumber\\
          &\approx& \frac{3u_{\BF}N^2}{4\pi R_{\rm TF}^3},
\end{eqnarray}
where we have used $n_{\B}\approx n_{\F}\approx N/(4\pi R_{\rm TF}^3/3)$ 
\cite{Akdeniz} with $R_{\rm TF} = \sqrt{2(6N)^{1/3}}$ for the system 
considered in the present work. As $a_\BF$ is increased further the system 
enters the partial demixing regime and the characteristic signature of which 
is a maxima in inter-species interaction energy. For our present calculations, 
the variation of the inter-species interaction energy with $a_\BF$ is shown in 
Fig. \ref{eint_bf}.  The condition for attaining partial demixing in spherical 
traps is \cite{Akdeniz} 
\begin{equation}
   a_\BF\ge\left(c_1\frac{N_\F^{1/2}}{N_\B^{2/5}}+c_2
           \frac{N_\B^{2/5}}{N_\F^{1/3}}\right)a_\BB,
\label{partial_demixing}
\end{equation}
 where 
\begin{equation}
  c_1 = \frac{15^{3/5}}{48^{1/2}}\frac{m_\F ^{3/2}}{2m_R m_\B ^{1/2}}
        a_\BB ^{3/5}
\end{equation}
and
\begin{equation}
  c_2 = \frac{48^{1/3}}{15^{3/5}}\left(\frac{6}{\pi}\right)^{2/3}
        \frac{m_\B}{2m_R}a_\BB^{2/5}
\end{equation}
The above condition for partial demixing is evaluated using TF approximation 
for the density profiles of both the components. From equation 
Eq.(\ref{partial_demixing}), the critical value of $a_\BF$ required to reach 
partial demixing regime for the Bose-Fermi mixture under consideration is 
$0.44a_\B$ and is significantly lower than the value of $0.7a_\BB$ obtained from
numerical solution of the coupled mean field equations Eq(\ref{eq:gpbf_r}). 
The difference may be attributed to the simplifying assumptions in deriving the
location of the inter-species interaction energy extrema. One of which is 
choosing the density profiles at $a_\BF=0$ to calculate the inter-species 
interaction energy. At $a_\BF=0.7a_\BB$ there is dramatic decrease in the 
density of the Bosons near the trap center. This is accompanied by 
corresponding decrease in overlap region between the two components as is 
shown in Fig.\ref{part_demixing}(b).

\begin{figure}[h]
   \includegraphics[width=8.5cm]{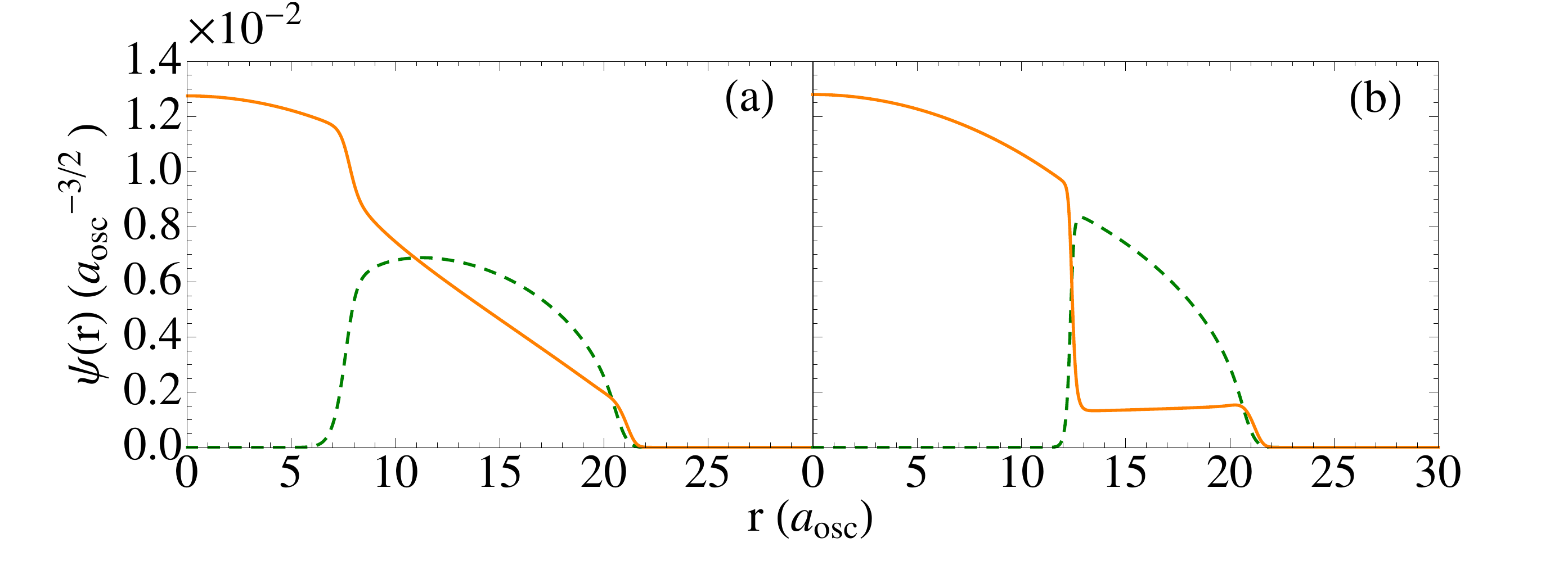}
   \caption{The wave function profiles of $^{174}$Yb (dashed-green line)
            and $^{173}$Yb (solid-orange line) in a mixture of 
            $^{174}$Yb-$^{173}$Yb, with $N_\B=N_\F=10^6$, $a_\BB=1100a_0$ and 
            $\omega/(2\pi)=400$Hz. For (a) $a_\BF=0.75a_\BB$ and for 
            (b) $a_\BF=1.0a_\BB$.}
\label{part_full_demixing}
\end{figure}

%%%%%%%%%%%%%%%%%%%%%%%%%%%%%%%%%%%%%%%%%%%%%%%%%%%%%%%%%%%%%%%%%%%%%%%%%%%%%%%
%%%%       Subsction III.B: Partial demixing to fully demixing regime     %%%%%
%%%%%%%%%%%%%%%%%%%%%%%%%%%%%%%%%%%%%%%%%%%%%%%%%%%%%%%%%%%%%%%%%%%%%%%%%%%%%%%

\subsection{Partial demixing to phase separation}

  A further increase of $a_\BF$, beyond the critical value, enhances the 
segregation of the two species. This lowers the inter-species overlap and 
balances the larger interaction energy from higher $a_\BF$. Ultimately, at 
higher values of $a_\BF$ the overlap is almost zero, the system can then be 
considered fully phase separated. The condition to attain phase separation
or fully demixed regime is \cite{Akdeniz}
\begin{equation}
  \alpha k_\F a_\BB > \left(\frac{a_\BB}{a_\BF}\right)^2,
\end{equation}
where 
\begin{equation}
  k_\F = (48 N_\F)^{1/6},\text{ and }
  \alpha = \frac{3^{1/3}}{4(2\pi)^{2/3}}\frac{m_\B m_\F}{m_R^2}.
\end{equation}
For $^{174}$Yb-$^{173}$Yb mixture with the previously mentioned parameters, 
the above criterion translates into $a_\BF>0.9a_\BB$. In the phase separated
domain, the separation occurs around the inner point where densities are equal.
To identify the location of this point consider the $a_\BF =0 $ density 
profiles. If the two profiles intersect at $ r_{\rm i}$, then from the TF 
approximation $r_{\rm i}$ is the solution of the equation
\begin{equation}
  \left\{2m_{\rm ratio} \left [E_\F - V_\F(r_{\rm i}) \right ]\right\}^3 =
  \left(\frac{6\pi ^2}{u_\BB} \right)^2\left[\mu - V_\B(r_{\rm i}) \right ]^2.
  \label{eqn_ri}
\end{equation}

For the system of our interest $V_\F$ and $V_\B$ are almost identical. 
Further, when $a_\BB$ is chosen (it satisfies Eq. (\ref{eqn_neql})) to match 
the spatial extents of the densities, $E_\F\approx \mu$. Following which, to 
a very good approximation
$\left [E_\F - V_\F(r_{\rm i})\right ]\approx \left[\mu - 
V_\B(r_{\rm i})\right ]$.   
The solution of Eq. (\ref{eqn_ri}) is then
\begin{equation}
  r_{\rm i} = \left [2E_\F -\frac{1}{4m_{\rm ratio}^3}
        \left ( \frac{6\pi ^2}{u_\BB}\right )^2 \right ] ^{1/2}.
\end{equation}

The importance of $r_{\rm i}$ is for the following: $n_\F(r) > n_\B(r)$ for 
$r<r_{\rm i}$, and $n_F(r) < n_\B(r)$ for $r>r_{\rm i}$. 
For the $^{174}$Yb-$^{173}$Yb mixture, based on the above relation 
$r_{\rm i}=13.09a_{\rm ho}$ for $N_\B=N_\F=10^6$ and $a_{\BB}=0.17a_{\rm ho}$, 
while the numerical value is $r_{\rm i}=12.42a_{\rm ho}$. 
Energetically, when $a_\BF$ is switched 
on it is favorable to accommodate the bosons and fermions at the outer and
inner regions about $r_{\rm i}$, respectively. As  $a_\BF$ is increased, the 
position of $r_{\rm i}$ tend to migrate outward but not dramatically. 

With further increase in $a_\BF$, the bosons are expelled towards the edge of
trapping potential while fermions are squeezed towards the trap center 
(see Fig.\ref{part_full_demixing}). With TF approximation, the effective 
potential experienced by $^{173}$Yb in the overlap region is 
\begin{eqnarray}
  V_{\rm eff} & = &\left(m_{\rm ratio} - \frac{g_\BF}{g_\BB}\right)
                   \frac{r^2}{2}\nonumber\\
        & \approx& \left(1 - \frac{g_\BF}{g_\BB}\right)\frac{r^2}{2},
  \label{eqn_platue}
\end{eqnarray}
where we have considered $m_{\rm ratio}\approx1$ for $^{174}$Yb-$^{173}$Yb 
mixture. Obviously, the effective potential experienced by fermions vanishes 
at $a_\BF=a_\BB$, and this explains the constant wave function profile of 
$^{173}$Yb in the overlap region as is shown in 
Fig.\ref{part_full_demixing}(b). Unlike two component BECs, the density of 
the fermions in the region occupied by bosons is not zero when the criterion 
for full demixing is satisfied.  

\begin{figure}[h]
  \includegraphics[width=8.5cm]{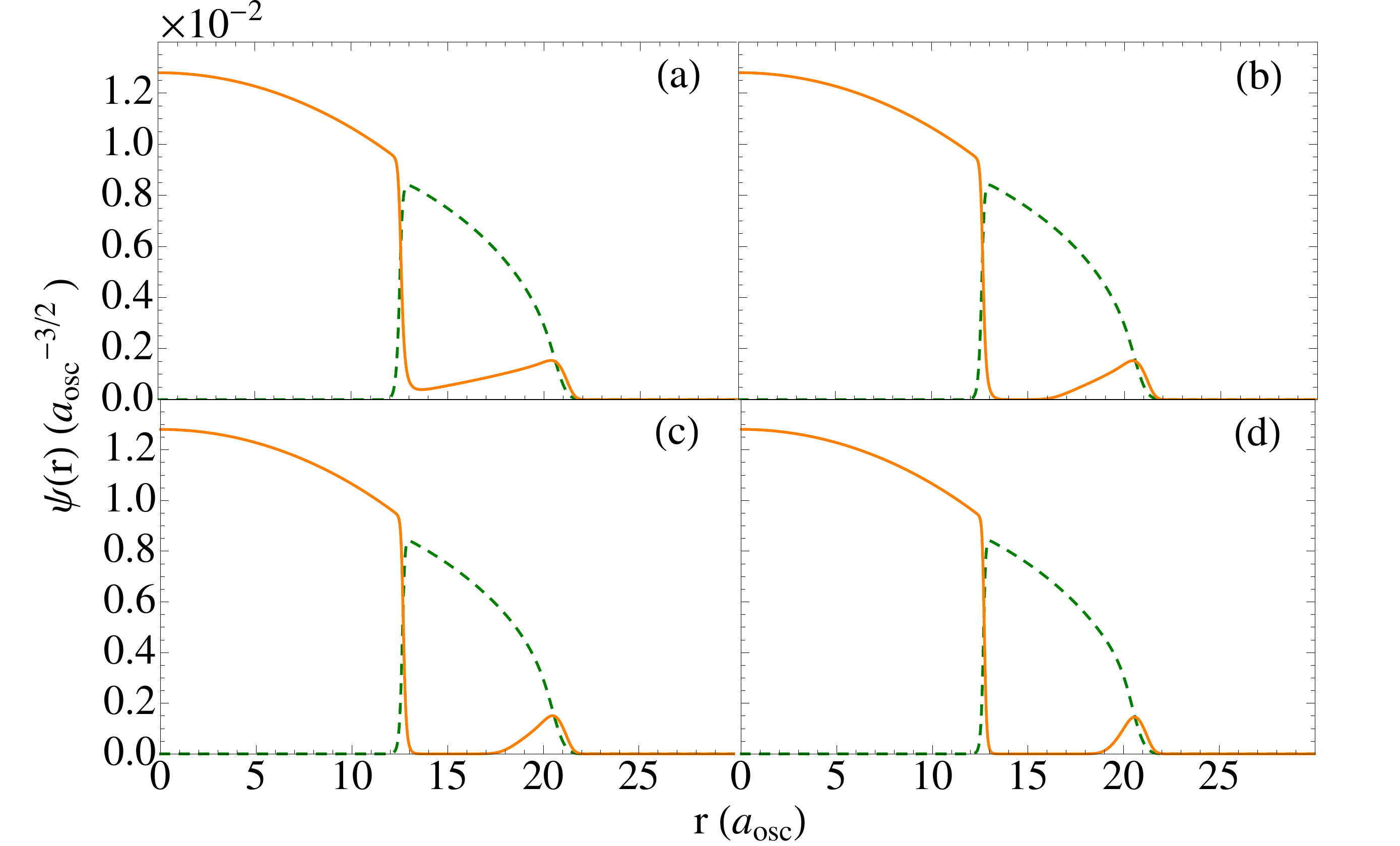}
  \caption{The wave function profiles of $^{174}$Yb (dashed-green line) 
           and $^{173}$Yb (solid-orange line) in a mixture of 
           $^{174}$Yb-$^{173}$Yb, with $N_\B=N_\F=10^6$, $a_\BB=1100a_0$ and 
           $\omega/(2\pi)=400$Hz, as $a_\BF$ is steadily increased from an
           initial value of $a_\BF=a_\BB$. For (a) $a_\BF=1.05a_\BB$,
           for (b) $a_\BF=1.1a_\BB$, for (c) $a_\BF=1.15a_\BB$, and for 
           (d) $a_\BF=1.25a_\BB$. }
\label{part_full_demixing_hump}
\end{figure}

%%%%%%%%%%%%%%%%%%%%%%%%%%%%%%%%%%%%%%%%%%%%%%%%%%%%%%%%%%%%%%%%%%%%%%%%%%%%%%%
%%%%                   Subsction III.C: Fermion pinching                  %%%%%
%%%%%%%%%%%%%%%%%%%%%%%%%%%%%%%%%%%%%%%%%%%%%%%%%%%%%%%%%%%%%%%%%%%%%%%%%%%%%%%

\subsection{Fermion pinching}

 For the values of $a_\BB $ marginally below $a^*_\BB$, besides $r_{\rm i} $ 
there is another point $r_{\rm o} $ where the densities are identical. The 
location of $r_{\rm o} $  is rather sensitive to kinetic energy corrections 
of the bosons \cite{Dalfovo}
\begin{equation}
   E_{\rm kin}=2.5\frac{N_{\B}}{R_{\rm TF}^2}\text{ln}
               \left(\frac{R_{\rm TF}}{1.3}\right).
\end{equation}
Without the kinetic energy correction, that is with TF approximation, 
$r_{\rm o }$ exists up to higher values of $a_\BF $. However, the kinetic
energy correction softens the profile at the edges and $r_{\rm o}$ vanishes
as $a_\BB$ approach $a^*_\BB$. In the phase separated domain when $r_{\rm o}$ 
is close to the edge, the fermion density is depleted at 
$ r_{\rm i} <r< r_{\rm o}$ for higher $a_\BF$. And, there is fermion density 
enhancement at $ r < r_{\rm i}$ and $r>r_{\rm o}$. For the bosons it is 
opposite: there is density enhancement at $ r_{\rm i} <r< r_{\rm o}$, and 
depletion at $ r < r_{\rm i}$ and $r>r_{\rm o}$.

As $a_\BF$ is increased to values larger than $a_\BB$, the effective 
potential within the overlap region ($ r_{\rm i} <r< r_{\rm o}$) is 
approximately
\begin{equation}
  V_{\rm eff} \approx \mu\frac{g_\BF}{g_\BB}- \eta\frac{r^2}{2},
  \label{eqn_pinch}
\end{equation}
where $\eta = |m_{\rm ratio} - g_\BF/g_\BB|$ and like in the previous case
we can take $m_{\rm ratio} \approx 1$. The form of $V_{\rm eff}$ is
repulsive with a maxima at $r_{\rm i}$ and decreases towards $r_{\rm o}$. 
The net effect is, the fermion density profile is pinched at the region
where $r$ is marginally larger than $r_{\rm i}$. Onset of pinching is
clearly discernible in Fig. \ref{part_full_demixing_hump}(a) and 
Fig. \ref{part_full_demixing_hump}(b) shows density profile with even higher
pinching effect. At higher values of $a_\BF$ the pinching is complete and 
an island of fermions appears at the edge. The
Fig. \ref{part_full_demixing_hump}(c-d) show the formation of the fermionic
island due to pinching in the $^{173}$Yb-$^{174}$Yb mixture considered in the
present work.

%%%%%%%%%%%%%%%%%%%%%%%%%%%%%%%%%%%%%%%%%%%%%%%%%%%%%%%%%%%%%%%%%%%%%%%%%%%%%%%
%%%%                  Section IV: Profile swapping                        %%%%%
%%%%%%%%%%%%%%%%%%%%%%%%%%%%%%%%%%%%%%%%%%%%%%%%%%%%%%%%%%%%%%%%%%%%%%%%%%%%%%%

\section{Profile swapping}

\begin{figure}[h]
\includegraphics[width=8.cm]{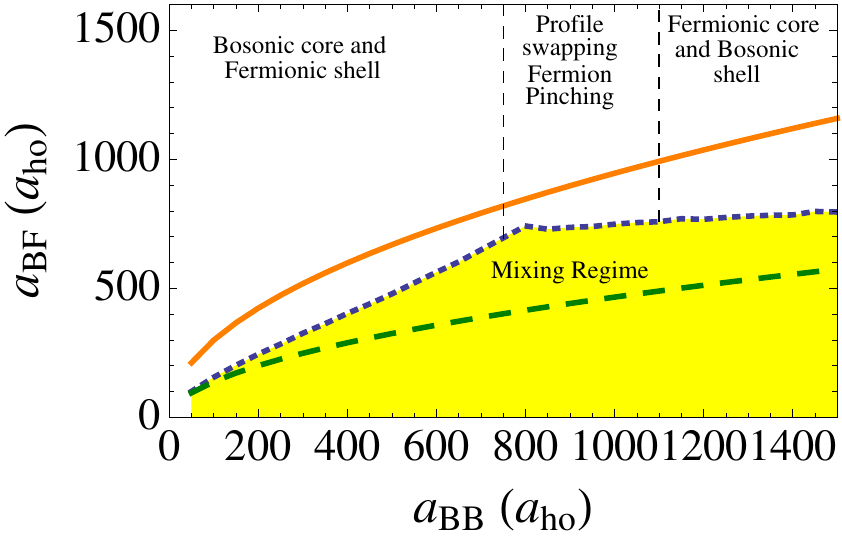}
\caption{The phase diagram of $^{174}$Yb-$^{173}$Yb mixture with 
$N_\B=N_\F=10^6$ and $\omega/(2\pi)=400$Hz. Dashed-green and solid-orange 
curves are the semi-analytic conditions for mixing to partial demixing 
and demixing to phase separation (or full demixing) transitions, while 
dotted-blue is the numerically obtained criterion for mixing to partial 
demixing transition. In phase separated regime, for $a_{\BB}\lesssim750$
bosonic core is surrounded by fermionic shell, for 
$750\lesssim a_{\BB}\lesssim1100$ there is fermion pinching along with
profile swapping at $a_{\BB}\approx750$. For $750\lesssim a_{\BB}\lesssim1100$,
bosonic shell is surrounded by fermionic core and shell. For $a_{\BB}\gtrsim1100$,
fermionic core is surrounded by bosonic shell.}
\label{phase_diag}
\end{figure}

 A remarkable feature in the evolution of density profiles as a function of 
$a_\BF$ is the observation of profile swapping for certain range of parameters.
In which the fermions are initially at the core and bosons form
a shell. However, at higher values of $a_\BF$ the bosons occupy the core and 
fermions forms a shell around it.
\begin{figure}[h]
  \includegraphics[width=8.5cm]{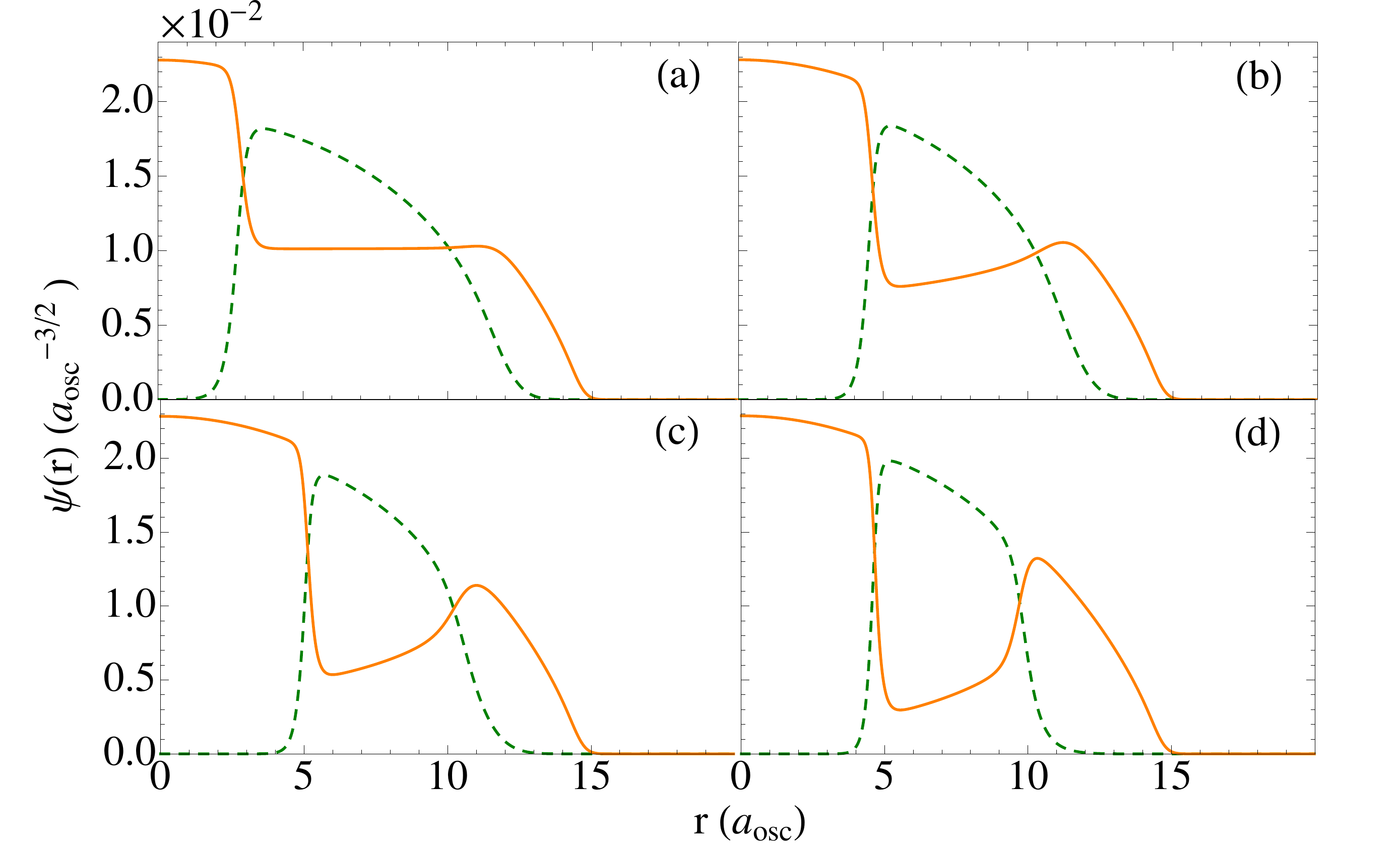}
  \caption{The wave function profiles of $^{174}$Yb (dashed-green line) 
           and $^{173}$Yb (solid-orange line) in a mixture of 
           $^{174}$Yb-$^{173}$Yb, with $N_\B=N_\F=10^5$, $a_\BB=1100a_0$ and 
           $\omega/(2\pi)=400$Hz, as $a_\BF$ is steadily increased from an
           initial value of $a_\BF=a_\BB$. For (a) $a_\BF=1.0a_\BB$, for 
           (b) $a_\BF=1.05a_\BB$, for (c) $a_\BF=1.1a_\BB$, and for 
           (d) $a_\BF=1.15a_\BB$.}
\label{N1.N2.100K_1}
\end{figure}
As an example to illustrate profile swapping, consider $N_B=N_F=10^5$, from 
Eq. (\ref{eqn_neql}) the spatial extents are equal at 
$a_\BB = 0.24 a_{\rm ho}$. However, retain the value $a_\BB=1100a_o $ as in 
the case of $10^6$ atoms in each species. In this case, the spatial extent of 
the bosons is less than the fermions, however, there are two points at which 
the densities of the bosons and fermions are the same. 
As mentioned earlier the ground state geometries of Bose-Fermi mixtures 
in spherical symmetric traps can be broadly categorized in three types, and
this is evident from the Fig.\ref{phase_diag} for $^{174}$Yb-$^{173}$Yb 
mixture with with $N_\B=N_\F=10^6$ and $\omega/(2\pi)=400$Hz. 

\begin{figure}[h]
\includegraphics[width=8.5cm]{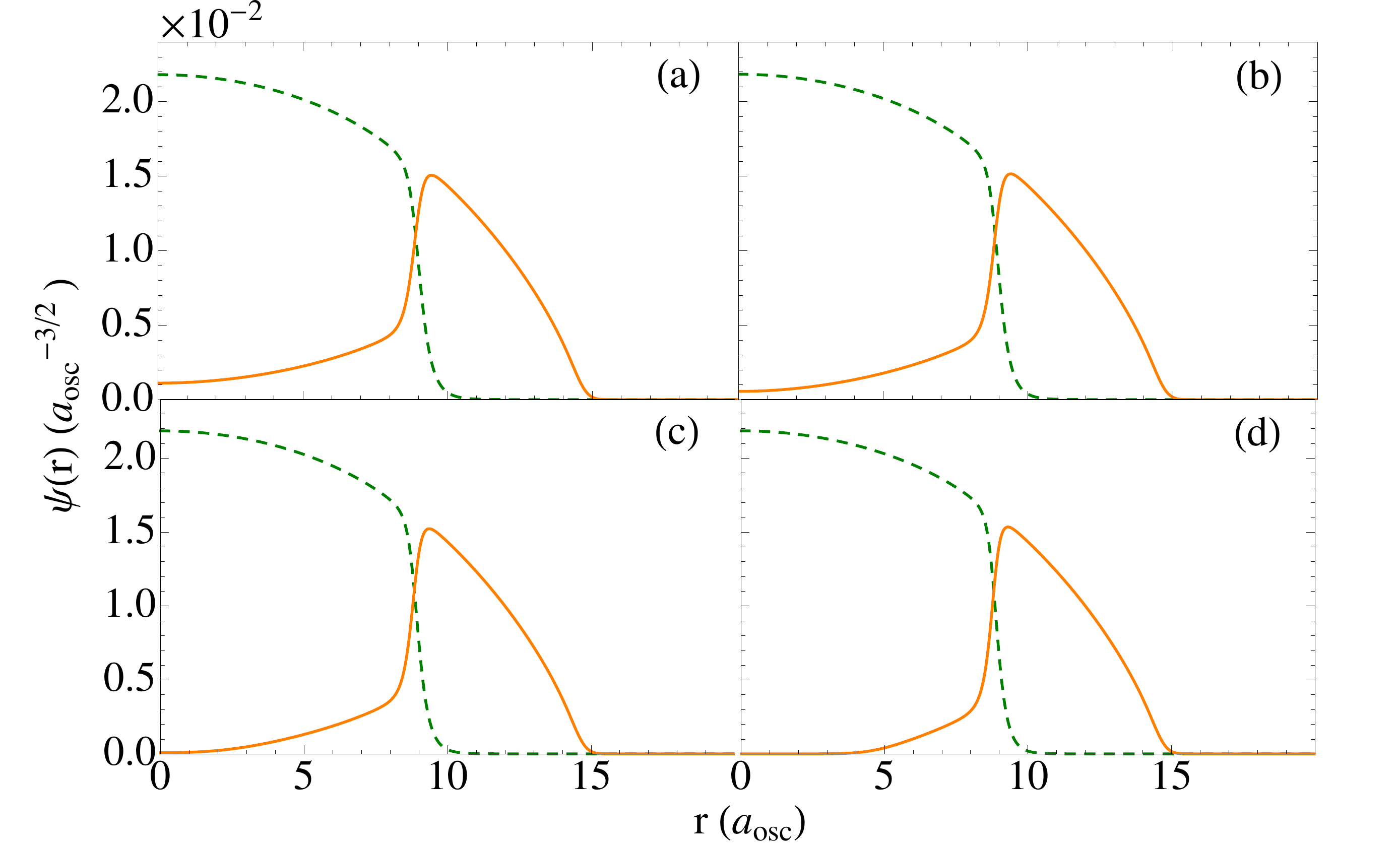}
\caption{The wave function profiles of $^{174}$Yb (dashed-green line)
         and $^{173}$Yb (solid-orange line) in a mixture of 
         $^{174}$Yb-$^{173}$Yb, with $N_\B=N_\F=10^5$, $a_\BB=1100a_0$ and 
         $\omega/(2\pi)=400$Hz, as $a_\BF$ is steadily increased from its 
         initial value of $a_\BF=1.16a_\BB$. For (a) $a_\BF=1.16a_\BB$,
         for (b) $a_\BF=1.17a_\BB$, for (c) $a_\BF=1.18a_\BB$, and for 
         (d) $a_\BF=1.2a_\BB$.}
\label{N1.N2.100K_2}
\end{figure}

 When $a_\BF$ is set to a non-zero value, at lower values the changes in the 
equilibrium density profiles exhibit a pattern  similar to fermion 
{\em pinching}.  Like in fermion {\em pinching } as $a_\BF$ is ramped up, 
there is a depletion of fermions from the overlap region as shown in 
Fig. \ref{N1.N2.100K_1}(a-d). However, at some value of $a_\BF$ a dramatic 
departure occurs. The fermions from the core are expelled to the edges and 
bosons settle at the core, Fig \ref{N1.N2.100K_2}. At intermediate values
of $a_\BF$, the bosons form a shell sandwich between fermions at the core
and an outer shell. This is evident from the density profiles shown in 
Fig. \ref{N1.N2.100K_1}(d).  As is evident from the figures, the migration of 
the fermions to the flanks occurs at a relatively minute change in $a_\BF$, 
from 1.15$a_\BB$ to 1.16$a_\BB$.

To analyze the profile swapping based on total energy considerations, take the 
density profiles just prior to the expulsion of fermions from the core. The
inter-species interaction energy is 
\begin{equation}
   E_{\rm int} \approx \int_{r_{\rm i}-\delta}^{r_{\rm i}+\delta}d\mathbf{r}
            u_\BF n_\B n_\F + \int_{r_{\rm i}+\delta}^{\infty}d\mathbf{r}
            u_\BF n_\B n_\F .
\end{equation}
Here, $r_{\rm i}$ like defined earlier is the inner point where the two 
densities are equal. The first term is the inter-species interaction energy 
arising from the inner boundary of the overlap region. And $\delta $ is the 
interpenetration depth considered symmetric for simplicity. The second term is 
the interaction energy from the remaining overlap region, though the upper 
limit of integration is taken as $\infty $ in reality it extends up to the 
point where $n_\B$ is nonzero. To simplify the analysis assume that the 
fermions from the core, after the position swapping, are pushed beyond the 
overlap domain. The interaction energy when swapping occurs is 
\begin{equation}
   E_{\rm int} \approx \int_0^{r_{\rm i}+\delta}d\mathbf{r}
            u_\BF n_\B n_\F + \int_{r_{\rm i}+\delta}^{\infty}d\mathbf{r}
            u_\BF n_\B n_\F ,
\end{equation}
where in the first term, the lower limit accounts for the nonzero fermion 
density around the core. The occurance of position swapping implies that
\begin{equation}
   \int_{r_{\rm i}-\delta}^{r_{\rm i}+\delta}d\mathbf{r} u_\BF n_\B n_\F 
   > \int_0^{r_{\rm i}+\delta}d\mathbf{r}
            u_\BF n_\B n_\F ,
\end{equation}
at some value of $a_\BF$. In other words, like in binary mixtures of 
condensates, at some point the geometry of the overlap region determines
the nature of the density profile. 

 Profile swapping, initiated by tuning interspecies scattering
length, appears to be a promising tool to study Rayleigh-Taylor type of 
instability in Bose-Fermi mixtures, which has been theoretically studied
in two species Bose-Einstein \cite{Gautam,Sasaki} condensates. The idea
is to start with a ground state geometry with fermions forming the core,
and then increase $a_{\BF}$ so that the new ground state has the fermionic
core swapped by bosonic one.

%%%%%%%%%%%%%%%%%%%%%%%%%%%%%%%%%%%%%%%%%%%%%%%%%%%%%%%%%%%%%%%%%%%%%%%%%%%%%%%
%%%%                  Section V: Conclusion                               %%%%%
%%%%%%%%%%%%%%%%%%%%%%%%%%%%%%%%%%%%%%%%%%%%%%%%%%%%%%%%%%%%%%%%%%%%%%%%%%%%%%%
\section{Conclusions}

  We have analyzed the equilibrium density profiles of the 
$^{174}$Yb-$^{173}$Yb Bose-Fermi mixture for a range of interaction strengths.
In this Bose-Fermi mixture, it is possible to tune both the Bose-Bose and 
Bose-Fermi interactions across a range of values. Density profiles of the
two species display pinching and position swapping when the boson-boson
scattering length is close to $a^*_\BB$, the value at which the spatial extent
of the bosons is same as the fermions. Pinching occurs when the $a_\BB$ is 
marginally below $a^*_\BB$, at these values as $a_\BF$ is increased fermions
at the edges are pinched to form a thin shell, whereas at even lower values of 
$a_\BB$, as $a_\BF$ is increased the fermions are expelled to the edge and 
density profiles are swapped. At intermediate values of $a_\BF$ the profiles
undergo through a series of configurations, and these are significantly different from
the ones in Bose-Bose mixtures. Close to the profile swapping domain, it 
should be possible to initiate Rayleigh-Taylor instability through a 
controlled variation of $a_\BF$. This would be significantly different from 
Rayleigh-Taylor instability in condensates. In future, it would be 
interesting and important to explore various instabilities which may occur
at the Bose-Fermi interface boundaries. These could be qualitatively different
from their analogues of binary condensates.

%%%%%%%%%%%%%%%%%%%%%%%%%%%%%%%%%%%%%%%%%%%%%%%%%%%%%%%%%%%%%%%%%%%%%%%%%%%%%%%
%%%%                      Acknowledgements                                %%%%%
%%%%%%%%%%%%%%%%%%%%%%%%%%%%%%%%%%%%%%%%%%%%%%%%%%%%%%%%%%%%%%%%%%%%%%%%%%%%%%%

\begin{acknowledgements}
We thank S. A. Silotri, B. K. Mani, and S. Chattopadhyay for very useful 
discussions. The numerical computations reported in the paper were done on
the 3 TFLOPs cluster at PRL.

\end{acknowledgements}

%%%%%%%%%%%%%%%%%%%%%%%%%%%%%%%%%%%%%%%%%%%%%%%%%%%%%%%%%%%%%%%%%%%%%%%%%%%%%%%
%%%%                        Bibliography                                  %%%%%
%%%%%%%%%%%%%%%%%%%%%%%%%%%%%%%%%%%%%%%%%%%%%%%%%%%%%%%%%%%%%%%%%%%%%%%%%%%%%%%

\end{document}